# Combinatorial Survey of Structural Phase Distribution and Magnetism in Fe-Ge-Te Composition-spread Thin Film Libraries

Chih-Yu Lee[1], Takahiro Yamazaki[2], Peng Yan[3], Ryan Kim[1], Masato Kotsugi[2], Efrain E. Rodriguez[4,5], Joseph W. Bennett[3], Ichiro Takeuchi[1,5*]

[1] Department of Materials Science and Engineering, University of Maryland, College Park, Maryland, USA
[2] Department of Materials Science and Engineering, Tokyo University of Science, Tokyo, Japan
[3] Department of Chemistry & Biochemistry, University of Maryland Baltimore County, Baltimore, USA
[4] Department of Chemistry and Biochemistry, University of Maryland, College Park, Maryland, USA
[5] Maryland Quantum Materials Center, University of Maryland, College Park, Maryland, USA
*Corresponding author

## Abstract

Recently, magnetic 2-dimensional (2D) van der Waals (vdW) materials have garnered tremendous attention. The vdW ferromagnet $Fe_5Ge_1Te_2$ has a Curie temperature $T_c$ of ≈ 270 K, which is tailorable by tuning the stoichiometry and the Fe deficiency to reach room temperature. To explore the expanded compositional space, we implemented combinatorial synthesis and high-throughput characterization to investigate the structural phase distribution and ferromagnetism of a Fe-Ge-Te thin film library. The library was prepared by magnetron co-sputtering followed by annealing in vacuum or in an inert environment. Composition and structural phase distribution of the 177 pads in the library were characterized using high-throughput wavelength dispersive spectroscopy (WDS), X-ray diffraction (XRD), and two-point probe resistance measurements. We leverage unsupervised machine learning to cluster the XRD dataset into groups of compositions with similar structural phases, and further study the ferromagnetic properties via SQUID magnetometry and X-ray magnetic circular dichroism (XMCD) across different clusters. The results are compared against magnetization and structural models calculated using DFT. Our results demonstrate that the hexagonal crystal structure is a critical prerequisite for ferromagnetism in this system, and that unexplored materials adopting this structure can be efficiently identified as possible ferromagnetic materials using our high-throughput, ML-assisted framework. This workflow based on the combinatorial strategy allows us to rapidly capture the composition-structure-magnetic property map across a broad compositional landscape of novel magnetic materials.

## 1. Introduction

Since the discovery of graphene, two-dimensional (2D) van der Waals (vdW) materials have become the focus of intense research as they provide a versatile platform for exploring magnetism in the monolayer limit.[1] The 2D vdW magnets, such as $Cr_2Ge_2Te_6$, $MnBi_2Te_4$, $FePS_3$, $CrI_3$ etc, exhibit exotic quantum phenomena and have great potentials in ultracompact spintronic applications.[2,3] Among these, the $Fe_xGeTe_2$ ($3 \leq x \leq 7$) family (FGT) is particularly promising owing to its broad spectrum of remarkable magnetic phenomena, including the anomalous Hall effect, robust tunneling magnetoresistance, strong out-of-plane magneto crystalline anisotropy, magnetic skyrmions, and magnetocaloric effect.[4–9] FGT as a metallic crystal preserves ferromagnetic order down to atomic layers and high Curie temperatures ($T_c$), which can be flexibly tuned through tuning Fe stoichiometry, strain engineering, hydrostatic pressure, optical illumination, electrostatic gating, proximity effect, elemental doping, electron/hole doping, intercalation, and patterning. [10–19] For example, $T_c$ of pristine $Fe_3Ge_1Te_2$ thin flakes is suppressed, relative to the bulk $T_c$ of 205K. With ionic liquid gating, however, the $T_c$ can be raised to 300 K.[20] As a result, a plethora of emerging devices are being made based on FGT, such as tunneling spin valves, magnetic tunnel junctions, and spin-orbit torque devices.[20–22]

The FGT crystals' layered structures allow for various stoichiometric variations where multi-atomic Fe-Ge slabs ($\approx$ 1 nm) are separated by vdW-bonded Te atoms. The layered nature leads to strong magnetic anisotropy with the easy axis often along the crystallographic c-axis, i.e., perpendicular to the layers. Fe atoms in Fe-Ge layers have inequivalent atomic sites with different valance states, i.e., $Fe^{3+}$ and $Fe^{2+}$, which govern the ferromagnetic structure, i.e. magnetic anisotropy, saturation magnetization, structural stability, and the hidden antiferromagnetic ordering that underpins the enhancement in $T_c$.[23] In addition, Fe vacancy or interstitial disorder, e.g., $Fe_{3-x}GeTe_2$ can lead to deviation in stoichiometry, resulting in modified crystal structure and hence magnetism. The lattice constant, especially the *c*-axis lattice parameter, can change slightly depending on Fe vacancies. A decrease in the Fe content leads to reduced $T_c$, smaller saturation magnetization, and in some cases, competing antiferromagnetic interactions or domain-wall pinning effects.[24] Additionally, Fe atoms hybridize with Te, which generate strong magnetic anisotropy because Te has robust spin-orbit coupling (SOC).[25] Te contributes not only to the magnetic anisotropy and to other spin-orbit phenomena, e.g., anomalous Hall effect, it facilitates controlling the interlayer spacing and the electronic overlap between slabs. The effect of the Te substitution at Ge sites and Fe-intercalation was investigated and was found to lead to the formation of local antiferromagnetic coupling.[26] The coexistence of ferromagnetic and antiferromagnetic phases in Te-rich FGT may potentially give rise to a spin-glass.[27] Ge occupies a central site within the Fe network as a stabilizer; its presence affects metallicity and hence can lead to a carrier-mediated magnetism.

It is important to note that due to the confluence of these chemical substitution effects, the magnetic ground state of $Fe_{5-x}GeTe_2$ is extremely complex, resulting in differing experimentally reported behaviors. Zhang *et al.* reported that $Fe_{5-x}GeTe_2$ undergoes two magnetic phase transitions as temperature decreases, including the ferromagnet to ferrimagnet at 275 K and glassy clusters below 110 K.[28] Another work has reported that $Fe_{5-x}GeTe_2$ transforms from the paramagnetic to ferromagnetic state at 265 K and then evolves into a ferrimagnetic state at 100 K.[29] These findings call for further systematic and broad chemical substitution investigations. On the computational side, Zhao *et al.* employed high-throughput first-principles calculations to systematically explore two-dimensional ternary transition-metal tellurides of the form X–Ge–Te (X = Fe, Mn, Cr), revealing a wide range of stoichiometries and geometric configurations with diverse magnetic properties.[30] On the experimental end, we carry out composition spread studies using a high-throughput approach, so that phase evolution and concomitant magnetic property variation can be tracked together as a function of continuously varying composition.

To date, the main technique for fabricating layered FGT films is chemical vapor transport (CVT)[31] followed by exfoliation via the Scotch-tape method, which often precludes systematic characterization across different compositions. Despite some progress which has been made through molecular beam epitaxy (MBE)[32,33] and pulsed laser deposition (PLD)[34], implementing large-area synthesis of high crystalline quality films, where the compositional range that can be explored in single experiments is limited. Bulk synthesis is also considered a low-throughput approach. Most importantly, due to the rare intra-layer three-dimensional network, magnetic behaviors in FGT are significantly different between bulk and layered counterparts. For example, the $T_c$ in monolayers is much lower than the bulk counterparts since magnetic ordering is sensitive to the thermal fluctuation in 2D structures.[35]

To address these challenges, we leverage co-sputtering to synthesize wide composition-range thin films in the form of composition spread libraries with a thickness equivalent to several hundred of atomic layers. Co-sputtered FGT films on $Si/SiO_2$ substrates are CMOS compatible, and thus once compositions of promising materials are identified, they can be readily scaled up for device fabrication. Zhao *et al.* investigated ultra-thin $Fe_3Ge_1Te_2$ in the amorphous state using magnetron sputtering and compared its magnetic properties to crystalline counterparts. [36] Because functional properties are intimately linked to structural properties, we first perform phase mapping across the compositional library. Herein, we employed wavelength dispersive spectroscopy (WDS) and x-ray diffraction (XRD) to collect compositional/ structural information in a high-throughput manner, and unsupervised machine learning is applied on all the diffraction patterns together to cluster the patterns into groups of compositions with similar structures. Once the structural phase boundaries are mapped across the compositional phase diagram, we have investigated the magnetic properties of the FGT films via SQUID magnetometry measurements (MPMS) and X-ray magnetic circular dichroism (XMCD) to delineate the composition–structure–property relationship. Such a workflow is broadly applicable to investigation of various functional materials at large as summarized in **Figure 1**. We have also carried out density functional theory (DFT) calculations to obtain computational insight on the electronic and magnetic properties, which allowed us to compare the predicted structures with experimental results.

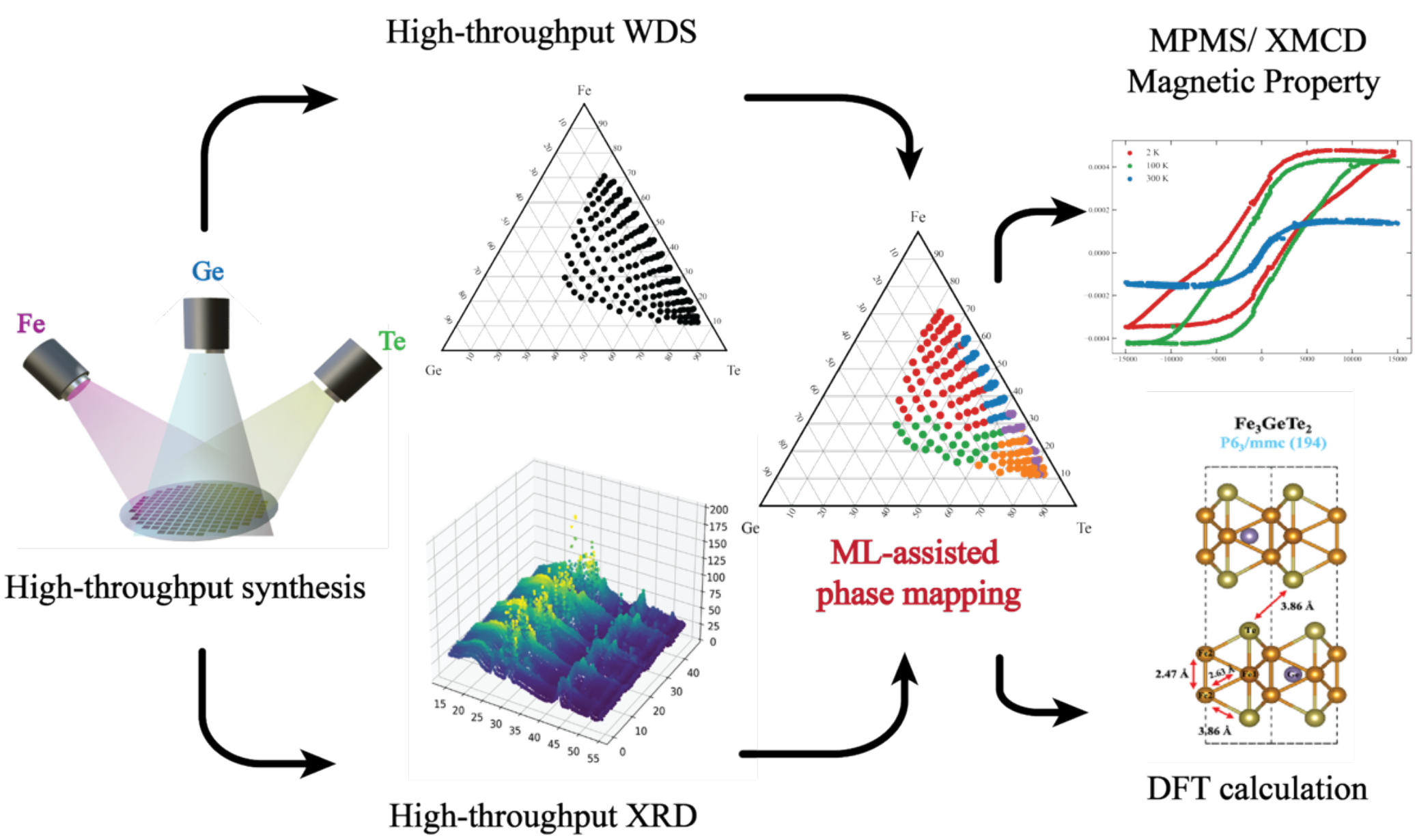

*Figure 1 The summary of Fe-Ge-Te composition-spread library study workflow from data collection to analysis. WDS and XRD characterization are high-throughput, while magnetic property measurement (MPMS/ XMCD) and DFT calculations are performed on selected individual compositions. With ML-assisted structural phase mapping, we can down select potentially ferromagnetic regions.*

## 2. Experimental Section

### 2.1 Thin film Synthesis

Ternary thin film composition spreads were fabricated on a $SiO_2$/Si substrate (thermal oxide: 2 μm) in an ultrahigh vacuum (base pressure: 2 x $10^{-8}$ Torr) magnetron sputtering system (AJA Orion-3) at room temperature. A patterned stainless-steel mask was placed on the substrate to define 177 gridded square regions (each 4 mm x 4 mm) separated by 0.5 mm gaps. High-purity Fe (99.995%), Ge (99.995%) and Te (99.995%) targets (1.5 in. diameter, 0.125 in. thick, Lesker) were co-sputtered with ultrahigh purity Argon gas (99.9997%, Airgas) with the pressure of 4.6 × $10^{-3}$ Torr. The thin film library was deposited over a period of 30 min using DC power sources for Fe at 35 W and RF power for Ge and Te at 23 W and 17 W, respectively. Note that co-sputtering results in intrinsic thickness gradient (90 nm-120 nm), confirmed by a profilometer. Multiple nominally identical composition spread samples were sputtered. The film was annealed through rapid thermal annealing (HEATPULSE 610) 20 mins for 350°C in $N_2$ environment.

### 2.2 Characterization of Basic Materials Properties

Chemical composition of the FGT thin film library was determined by using WDS in an electron probe microanalyzer (EPMA) JXA 8900R Microprobe, with an acceleration voltage of 15 kV. Calibration was done using polished pure metal with an experimental error margin of < 0.3 at. %. DISCOVER powder diffractometer (Bruker C2/D8) with a CuKα radiation source was used to collect XRD images. Diffraction data were collected using two frames with an exposure time of 5 minutes per frame, covering 2θ ranges of 16–35° and 35–54°, respectively, and subsequently integrated into 1D patterns over 16–54° in 2θ with a step size of 0.05°. For 177 separate composition regions, electrical sheet resistance was measured using an automated probe station (Signatone WL210) via digital multimeter (Keithley 2400) using 1mA. The results are an average value of 10 measurements/square composition region. The magnetization measurements were accomplished in a SQUID by Quantum Design with magnetic field up to 7 T. The zero-field-cooled (ZFC) and field-cooled (FC) measurements were done at 100 Oe. M vs. H hysteresis measurements were carried out from 0 to 1.5/-1.5 T. Magnetization was calculated by normalizing the measured magnetic moment by the film volume; out-of-plane measurements are shown without demagnetization correction.

### 2.3 Machine-learning Methods

A Python-based unsupervised machine-learning algorithm was used to cluster XRD and composition data. A combined similarity matrix was used (Cosine and Euclidean distance were used for structural and compositional similarity matrix). Spectral embedding was employed for non-linear dimensionality reduction.[37] A Gaussian mixture model was used to estimate the probability distribution for N (N = 5) clusters.[38] The choice of the number of clusters of 5 was selected based on the evaluation of clustering performance using the elbow method, which estimates how the within-cluster variance decreases as the number of clusters increases. (see **Supplementary S1**)

### 2.4 X-ray Magnetic Circular Dichroism (XMCD)

The XMCD measurements were performed at the BL25SU beamline of the SPring-8 synchrotron radiation facility in Japan. X-ray absorption spectroscopy (XAS) ($\mu^{+}+\mu^{-}$) and XMCD spectra ($\mu^{+}-\mu^{-}$) were obtained at 10K for both in-plane (0°) and out-of-plane (80°) directions. M-H hysteresis loops were collected from 0 to 1.91 T/-1.91 T at Fe $L_3$ edge (706.8 eV), while MCD spectra were measured across the Fe $L_2$ and $L_3$ edges (680 eV- 760 eV). Magneto-optical sum rules [39,40] were applied to data analysis with the assumption that the number of unoccupied 3d states is 3.39. Argon ion milling was employed prior to measurements to remove an oxidized surface layer, as XMCD is highly sensitive to surface conditions. Further details on XMCD measurements are available in **S2-S3** and Yamazaki *et al*.[41] The detailed analysis are shown in **S5-S10**.

### 2.5 DFT Calculations

First-principles calculations were conducted using the open-source software package QUANTUM ESPRESSO[42], which implements periodic plane-wave density functional theory (DFT)[43–45]. The Perdew–Burke–Ernzerhof (PBE)[46] functional within the generalized gradient approximation (GGA) framework was employed, along with the van der Waals interaction (vdW) correction using Grimme's DFT-D2 method[47] (with a global scaling factor $S_6 = 0.5$) for all calculations. GBRV[48] ultrasoft pseudopotentials were used to describe all atoms with a plane-wave basis cutoff of 40 Ry. Structural relaxations were carried out using a 6 × 6 × 6 Monkhorst–Pack k-point grid[49]. Self-consistent field (SCF) calculations were converged to an energy threshold of $1 \times 10^{-7}$ eV and a force criterion of 1–5 meV/Å per atom.

## 3. Results and Discussions

### 3.1 XRD

By annealing a spread at 350 °C under flowing $N_2$ at ambient pressure, the composition range covered on the library wafer shows minimal change from the as-deposited state to the crystalline state. The stoichiometry measured from an annealed spread is shown in **Figure 2-a**, covering compositions of well-known FGT materials, such as $Fe_3Ge_1Te_2$ and $Fe_5Ge_1Te_2$. It is worth noting that they are clustered structurally as belonging to the same group via unsupervised machine learning, which indicates they have a similar hexagonal crystal structure (space group: $P6_3/mmc$), and the diffraction patterns show a series of {002} planes— (004) and (006) and other related peaks.

We have focused our investigation on the composition range which spans $Fe_xGeTe_2$ (x = 1, 3 and 5). A clear structural phase transition is observed in going from x = 1 to x = 5. At x = 1, there are some impurity phases besides the main FGT structure, as seen in **Figure 2-b**, as $Fe_1Ge_1Te_2$ is also located at the phase boundary identified by ML-assisted phase mapping. When x = 5, the film is prone to be more crystalline, but diffraction along the c-axis is weakened. We surmise that the addition of Te to lead to existence of multiple phases. It is observed that Te-rich phases tend to crystallize more readily as Te crystallizes at 200 °C.[50] Higher temperatures can affect the morphology and associated thermodynamics. Another composition we looked at to further study the structural phase boundary is $Fe_2Ge_1Te_4$, which is a previously unexplored material. According to XRD patterns, it remains an FGT-like structure while containing a Te- rich phase. Electrical sheet resistance spans from 50 to 350 Ohm at the Ge-Te-rich edge (**Figure S4**) in the as-deposited state, while after annealing, the resistance falls between 50-150 Ohm. Resistance in the crystalline state doesn't change significantly across the library wafer and the small resistance variation is likely due to the thickness variation. It is known that the FGT family is metallic and considered a "Hund's metal".[51]

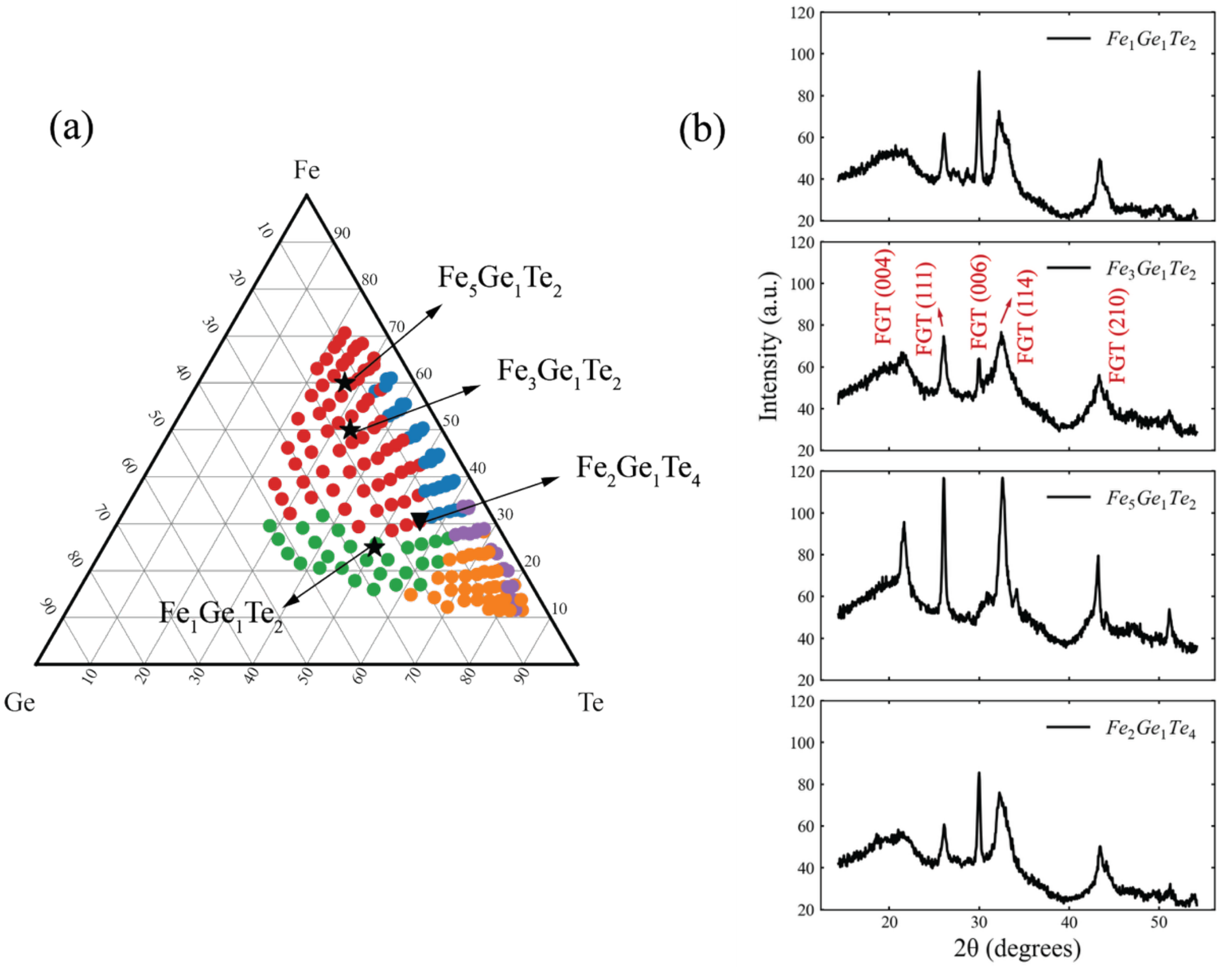

*Figure 2 ML-assisted structural phase mapping of XRD patterns. (a) Colors represent different clusters. Data points denoted by stars and the triangle are materials used to benchmark against previous reports. (b) Diffraction patterns of selected compositions.*

The ML-assisted phase mapping results are based on 5 distinguished phases, shown in **Figure 3-a**. The origin of ferromagnetism of the expanded FGT family is highly dependent on their crystal structures, and phase mapping can be exploited as a surrogate model to predict their magnetic properties. Group 1 consists of multiple Te-associated phases— FeTe hexagonal ($P6_3/mmc$), GeTe trigonal ($R3m$) and Te trigonal ($P3_121$). Group 2 are mostly a hexagonal FGT structure. Group 3 has both FGT and GeTe phases. Group 4, with minimal Ge content, has a small amount of FGT, and FeTe is the predominant phase, while Group 5 consists of FeTe and distinct Te phases. Note that sample labels correspond to their nominal compositional positions in the mapping. Although impurity phases may exist, the materials are designated by these composition-based names for simplicity.

3.2 SQUID

In general, materials properties including magnetism are associated specific structural phases. From the structural phase mapping discussed above, based on the known ferromagnetic compounds[25,36,52], we identify the red colored composition region in **Figure 2-a** (roughly covering $Fe_{1-x-y}Ge_xTe_y$, with approximately $0.02 \leq x \leq 0.57$ and $0.15 \leq y \leq 0.85$) to be a pool of potential novel ferromagnetic materials candidates. Two materials are selected for SQUID measurements — $Fe_5Ge_1Te_2$, which is a known material which can serve as a benchmark and previously unexplored $Fe_2Ge_1Te_4$. Since the easy axis of FGT films is reported to be along the c-axis[53], which is out-of-plane, the measurement was set up with magnetic field applied perpendicular to samples. **Figure 3-b** and **Figure 3-c** display the zero-field-cooled (ZFC) and field-cooled (FC) curves under 100 Oe. We find the $T_c$ of $Fe_5Ge_1Te_2$ is ≈ 235 K, while that of $Fe_2Ge_1Te_4$ is ≈ 155 K. The $T_c$ of $Fe_5Ge_1Te_2$ we deposited was found to be lower than that from exfoliated flakes from a bulk crystal, which is $T_c \approx 300$ K[54,55], and we attribute the discrepancy to the low crystallinity of our film sample. There is possible coexistence of short-range and long-range ferromagnetism in $Fe_5Ge_1Te_2$, which has been reported in a hybrid structure.[56] The structural phase belongs to, Group 2 in **Figure 3-a**, has overall low crystallinity, which implies insufficient annealing process for the films. We note that it is challenging to crystallize this region without destroying other volatile phases. 350 °C was selected as a trade-off optimal temperature for the entire spread. We see significant change in coercive field and saturation magnetization from RT (300 K) to 100 K in both films, which is consistent with magnetic transition observed in ZFC/FC measurements. $Fe_5Ge_1Te_2$ has higher magnetization saturation than that of $Fe_2Ge_1Te_4$, while its coercivity is surprisingly lower. It is noted that $Fe_2Ge_1Te_4$ exhibits weak ferromagnetism at room temperature, which might indicate the enhancement of ferromagnetism owing to Te-doping.[57] Due to the low-throughput nature of SQUID measurements, we did not measure all the compositions from the spread.

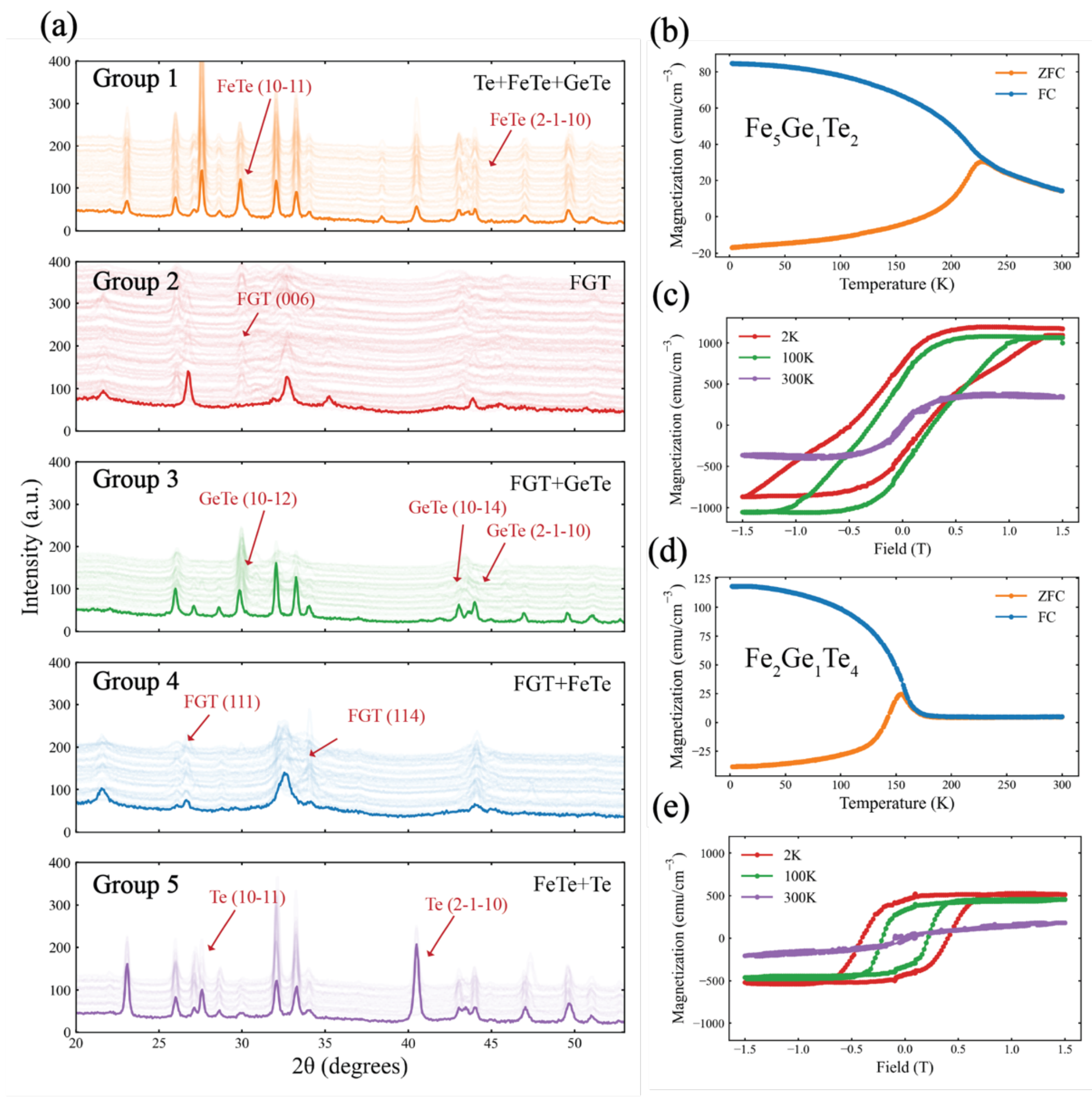

*Figure 3 Structural analysis and magnetic properties. (a) XRD analysis based on clustering results. The color in each subplot corresponds to Figure 2-a. Diffraction peaks are labeled for each phase. Note that some peaks are not identified. (b) and (d) are ZFC/ FC scans for $Fe_5Ge_1Te_2$ and for $Fe_2Ge_1Te_4$, respectively; (c) and (e) are M-H hysteresis loops for $Fe_5Ge_1Te_2$ and $F_2GeTe_4$, respectively at 2K, 100K and 300K.*

### 3.3 XMCD

To investigate the spin states of Fe sites with different valence states, XMCD measurements were employed. Based on the selection rule and the orbital nature, the Fe $L_{2,3}$ edges correspond to the 2p → 3d excitation process. We measured 3 compositions — $Fe_3Ge_1Te_2$ (Group 2, FGT), $Fe_{0.8}Ge_1Te_{1.8}$ (Group 3, FGT+ GeTe) and $Fe_2Ge_1Te_4$ (Group 2, FGT) to compare the effects of varying Fe concentration and the existence of impurity phase(s). Note that although $Fe_{0.8}Ge_1Te_{1.8}$ and $Fe_1Ge_1Te_2$ (sample measured in Section 3.2) have similar stoichiometries, they are treated as distinct samples; their assignment to the same group arises from similarities identified by the ML clustering rather than identical compositions. We find that in both in-plane and out-of-plane directions, $Fe_2Ge_1Te_4$ has the highest saturation magnetization, remanent magnetization and coercive field among the 3 samples, which is being considered for use as a hard magnet for spintronics (**Figure 4-a and 4-b**). Yet, the origin of enhanced coercivity is unclear as the materials consist of multiple phases, requiring further microstructure study. The hysteresis loops of Group 2 compositions ($Fe_3Ge_1Te_2$ and $Fe_2Ge_1Te_4$) display higher magnetization than $Fe_{0.8}Ge_1Te_{1.8}$, which suggests that the FGT hexagonal structure is the key to ferromagnetism.

The XMCD sum-rule analysis provides direct insight into the orbital moment anisotropy, which is closely linked to magnetocrystalline anisotropy. **Figure 4-c** shows MCD ($\mu^+$- $\mu^-$) from XAS for Fe $L_{2,3}$ and the completed post-processing is included in **Supplementary S11-S12**. We observe that $Fe_3Ge_1Te_2$ and $Fe_{0.8}Ge_1Te_{1.8}$ exhibit significantly larger orbital moments for in-plane magnetization compared to out-of-plane ($\Delta m_l$ = +0.15 and +0.25 μB, respectively, as shown in **Figure 4-g**), indicating strong orbital moment quenching when magnetization is forced perpendicular to the film. $Fe_2Ge_1Te_4$, in contrast, shows nearly identical orbital moments in both directions, as can be seen in **Figure 4-d**. Yet, the orbital moments extracted from our XMCD study are significantly larger than those from previous reports, i.e., 0.1-0.15 $\mu_B$ per Fe atom, especially in the in-plane direction. The overestimation can be attributed to several reasons, e.g., incorrect magnetic dipole term or number of 3d holes in sum rules in less symmetric systems[58], background subtraction/ normalization, strong SOC and interface effects in low dimensional layered structures[59] and unquenched orbital moments due to reduced coordination and symmetry breaking in 2D vdW magnets.[60] Yamagami et al. observed a large XMCD signal for the heavy element Te, which is an indication of a possible major role of SOC which affects both magnetocrystalline and anisotropic exchange interactions, for determining detailed arrangement of Fe derived spins.[61] Te has large intrinsic from 5p electrons, which strengthens the Fe-Te hybridization[53] or contribute through high Te px,y DOS near the Fermi level, leading to increased magnetic anisotropy[62]. Unfortunately, the limited beam time precluded us from investigation beyond the moments of Fe atoms. The corresponding spin moments, plotted in **Figure 4-e**, show comparable values for the two orientations (for $Fe_3Ge_1Te_2$ and $Fe_2Ge_1Te_4$) and suppressed out-of-plane spin moment for $Fe_{0.8}Ge_1Te_{1.8}$. The average spin moments are comparable to those reported in other literatures, i.e., 1.6-1.8 $\mu_B$ per Fe atom for $Fe_5Ge_1Te_2$[52,61] and 1-2 $\mu_B$ per Fe atom for $Fe_3Ge_1Te_2$[63,64], which vary with Fe sites and stoichiometry as Fe deficiency leads to reduction in both moments. Total moments ($\Delta m_{total} = m_l + m_s$) in **Figure 4-f** follow the same trend as spin moments.

The orbital moment anisotropy $\Delta m_l = m_l^{\parallel} - m_l^{\perp}$, plotted in **Figure 4-g**, is closely related to the SOC-driven magnetocrystalline anisotropy. $Fe_3Ge_1Te_2$ and $Fe_{0.8}Ge_1Te_{1.8}$ exhibit large $\Delta m_l$, demonstrating that the magnetocrystalline contribution in our films favors an in-plane orientation while $Fe_2Ge_1Te_4$ exhibits $\Delta m_l \approx 0$, indicating negligible magnetocrystalline anisotropy. This behavior contrasts with the out-of-plane anisotropy of bulk (or exfoliated flakes from bulk) FGT, reflecting the altered magnetic environment of our polycrystalline films, where strong in-plane texturing modifies the magnetic anisotropy. The deviation from bulk properties due to strain has also been observed in thin films in previous studies.[65,66] Additionally, strain, disorder, and grain-boundary environments modify the Fe–Te coordination and suppress the out-of-plane orbital unquenching expected in single crystals. As a result, the magnetocrystalline contribution is significantly reduced or even inverted. As a result, the shape anisotropy inherent to thin films may become dominant, enforcing an in-plane easy axis. For $Fe_3Ge_1Te_2$ and $Fe_{0.8}Ge_1Te_{1.8}$, the orbital anisotropy also favors in-plane magnetization, further strengthening this tendency. $Fe_2Ge_1Te_4$, lacking orbital anisotropy, is governed primarily by shape anisotropy. Bulk FGT single crystals display a strong c-axis easy axis due to substantial SOC and uniaxial crystal symmetry[67]. However, our polycrystalline FGT thin films exhibit a pronounced in-plane magnetic easy axis primarily due to the shape anisotropy. MCD provides a direct microscopic confirmation that the orbital and spin moment landscape of these films is fundamentally modified relative to their bulk counterparts. Despite limited beam time, the general trend suggests the impurity phase and Fe deficiency decrease the total magnetic moments as it was reported that both spin and orbital moments decreased with Fe deciciency.[63] $Fe_3Ge_1Te_2$ and $Fe_2Ge_1Te_4$, with fewer Fe vacancies, resulting in stronger magnetic response from the samples, while $Fe_{0.8}Ge_1Te_{1.8}$, with more Te content, exhibits large in-plane and out-of-plane anisotropy. However, Audehm et al. reported that the broken local symmetry can pin orbital moments, thus leading to high anisotropy.[68] Detailed investigation of individual thin films is under way to further clarify the origin of the magnetic moment and anisotropy.

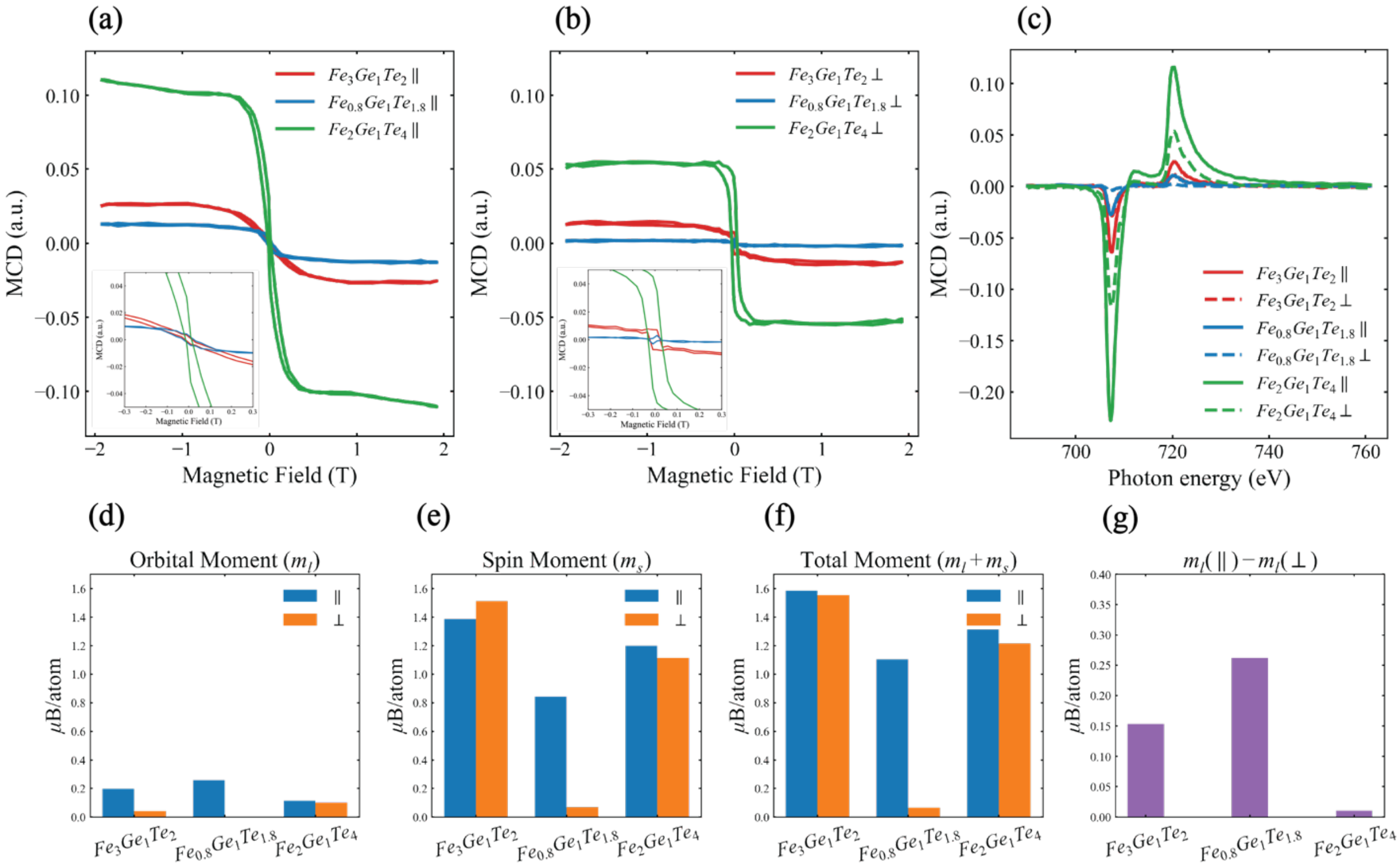

*Figure 4 XAS and XMCD for $Fe_3Ge_1Te_2$, $Fe_{0.8}Ge_1Te_{1.8}$ and $Fe_2Ge_1Te_4$. (a) M-H loop (hν = 706.9 eV) with both x-ray and magnetic field parallel to sample (0°). (b) M-H loop with both x-ray and magnetic field perpendicular to sample (80°). The insets in (a) and (b) are the magnified views in proximity of zero field. (c) is the XMCD ($\mu^+$- $\mu^-$) before normalization. Magnetic moments are displayed in (d) orbital moments $m_l$, (e) spin moments $m_s$, (f) total magnetic moments $m_l + m_s$ and (g) $\Delta m_l$, orbital moment difference between in-plane and out-of-plane direction.*

### 3.4 DFT

To gain theoretical insight into the experimental results, first-principles calculations are employed. There has been a report of good agreement between first-principles calculations and experiments on ferrimagnetic $Fe_3GaTe_2$[69], which shares similar structural and magnetic properties with $Fe_3Ge_1Te_2$. We adopted two approaches to construct the potential $Fe_2Ge_1Te_4$ structure, based on experimentally determined $Fe_3Ge_1Te_2$ (ICSD entry: 415616)[70] and $Fe_5Ge_1Te_2$ (ICSD entry: 130074)[71]. In both cases, we partially substituted Fe atoms on one Wyckoff sublattice with Te atoms and either added a Te layer or removed an extra Fe layer to maintain the stoichiometry of $Fe_2Ge_1Te_4$, with an exact atomic ratio Fe:Ge:Te of 2:1:4. Substitution details are described in the supplement for all three configurations. As we found in previous XRD characterizations, $Fe_2Ge_1Te_4$ inherits a similar hexagonal crystal structure as $Fe_3Ge_1Te_2$, and the most energetically-favorable configuration is shown in **Figure 5**. The other two configurations are included in **S13** (config-1) and **S14** (config-3). Based on DFT relaxation calculations, $Fe_2Ge_1Te_4$ breaks the Te–Te bridges within the cage and they separated the Te atoms from the main layer. The Te–Te distance increases from 2.66 Å in both input configurations (config-1 and config-2) to over 3.3 Å after relaxation. According to the bulk crystal structure of Te in the ICSD (entry: 96502)[72], the reference Te–Te bond length is approximately 2.79 Å. Therefore, this significant increase suggests a reluctance to form Te–Te dumbbells within the plane. However, the relaxed configuration indicates the possible formation of a freestanding Te–Te layer within the vdW gap between adjacent Fe–Ge–Te building blocks. The DFT-calculated total energy, average magnetic moment of Fe, total magnetization, lattice parameter and space group are summarized in Supplementary **S15**. Experimentally, we didn't observe any phase separation in XRD, which suggests the thin films we investigated are perhaps in a metastable state, or not fully crystalline. In other words, our results indicate that the experimentally derived compositions are more complex than the simple atomic lattice substitutions described here, though the similarities in magnetic moments point towards similarities in chemical environments that result from layering present in the structures. In the modeled compositions the separation of Te layers was 1:1 $Fe_2GeTe_2$:$Te_2$, in essence showing how strong the Fe-Ge interaction is when compared to the Ge-Te layering that resulted for multiple configurations. The pure $Fe_3Ge_1Te_2$ exhibits an average magnetic moment of 1.54 $\mu_B$ at the $Fe_1$ site (cage center) and 2.39 $\mu_B$ at the $Fe_2$ site (cage edge)[69]. The calculated $Fe_2Ge_1Te_4$ in configuration 2# shows similar magnetic moment of 1.87 $\mu_B$ and 1.93 $\mu_B$ for the two Fe sites (both at cage edges), due to their similar coordination environments. These values lie between those of the two distinct Fe sites in $Fe_3Ge_1Te_2$ but reflect a higher symmetry in magnetic properties (lower anisotropy), consistent with our experimental observations. Even though we are unable to calculate every compound in our library, in the future, the high-throughput structural analysis combined with ML-guided clustering, magnetic properties can be estimated through ab initio calculations.[73]

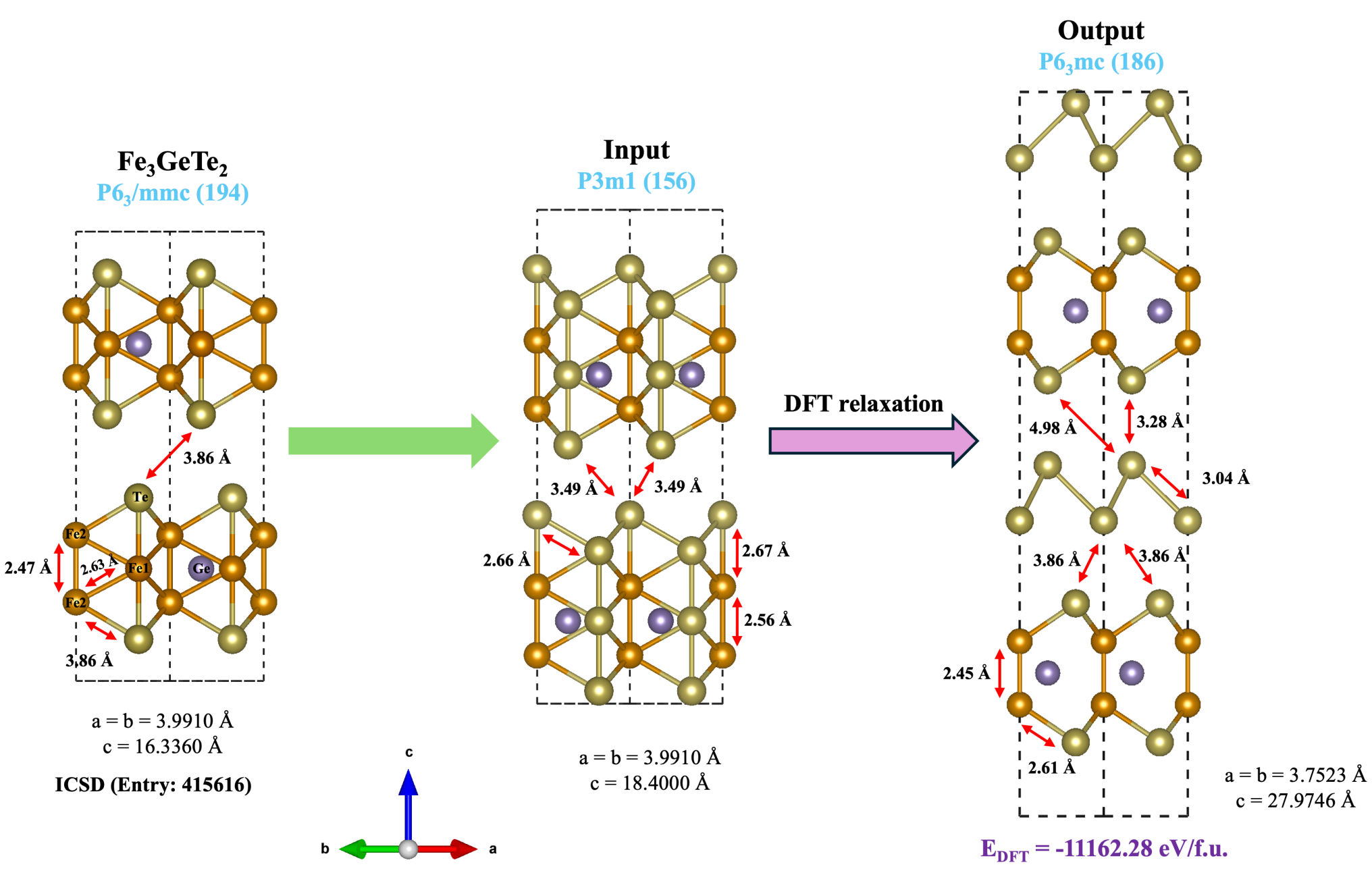

*Figure 5 DFT-simulated configuration 2# of $Fe_2Ge_1Te_4$ based on experimental data for $Fe_3Ge_1Te_2$.*

## 4. Conclusions

Our results demonstrate the successful growth of thin film Fe-Ge-Te via magnetron sputtering. We were able to synthesize several well-known ferromagnets within an expanded compositional space, despite the fact that the entire wafer had to be processed together, inevitably leading to potentially suboptimal processing conditions for some compositions. Unsupervised ML clustering of XRD patterns was used to obtain mapping of different crystal structure regions. By associating the crystal structure as the key factor determining magnetic properties, the clustering as allowed us to identify a potentially new ferromagnetic region in a large composition space. We investigated previously unexplored materials, e.g., $Fe_{0.8}Ge_1Te_{1.8}$ and $Fe_2Ge_1Te_4$, and compared them with well-known materials in the FGT family. We revealed the effect of the impurity phase and Fe deficiency on ferromagnetism via SQUID and XMCD. We were able to leverage DFT to provide microscopic insights on their potential magnetic behavior.

This work highlights the utility of the combinatorial approach to rapidly explore new magnetic materials by combining high-throughput characterization (composition mapping and XRD) with limited-throughput techniques (SQUID and XMCD) and chart composition-structure-property relationships in complex materials systems such as Fe-Ge-Te.

## Acknowledgement

This work is funded by DTRA through grant number HDTRA12410015. The XMCD experiments at BL25SU was carried out under the approval of SPring-8, Japan (Proposals: 2025B1250, 2025A1390, 2024B1320). Calculations were performed, in part, using the UMBC High Performance Computing Facility (HPCF), supported by the National Science Foundation under the MRI grants CNS-0821258, CNS-1228778, and OAC-1726023 and the SCREMS grant DMS-0821311. This research also used the Theory and Computation facility of the Center for Functional Nanomaterials (CFN), which is a U.S. Department of Energy Office of Science User Facility, at Brookhaven National Laboratory under Contract No. DE-SC0012704.